\documentclass[12pt]{iopart}
\usepackage{epsfig}
\begin{document}
\def\as{\alpha_{\mbox{\tiny S}}}
\def\alb{\bar\as}
\def\cG{{\cal G}}
\def\gE{\gamma_{\mbox{\tiny E}}}
\def\DL{{\mbox{\scriptsize DL}}}
\def\SL{{\mbox{\scriptsize SL}}}
\def\mR{\mu_{\mbox{\tiny R}}}
\def\res{\,{\mbox{\scriptsize jet}}}
\def\om{\omega}
\def\alom{\frac{\alb}{\om}}
\def\omal{\frac{\om}{\alb}}
\def\glip{\gamma_{\mbox{\tiny L}}}
\def\VEV#1{\overline{#1}}
\def\sig{\sigma}
\newcommand{\beq}{\begin{equation}}
\newcommand{\eeq}{\end{equation}}
\newcommand{\bea}{\begin{eqnarray}}
\newcommand{\eea}{\end{eqnarray}}
\begin{flushright}
DTP/99/114 \\
DAMTP-1999-179 \\
Cavendish-HEP-99/19
\end{flushright}
\boldmath
\title[BFKL Dynamics at Hadron Colliders]{BFKL Dynamics 
 at Hadron Colliders}
\unboldmath

\author{Carlo Ewerz$^{a,b,1}$, Lynne H.\ Orr$^{c,2}$, 
W.\ James Stirling$^{d,e,3}$ and Bryan R.\ Webber$^{a,f,4}$}

\address{$^a$ Cavendish Laboratory, Cambridge University,
Madingley Road, Cambridge CB3 0HE, UK
}
\address{$^b$ DAMTP, Centre for Mathematical Sciences, Cambridge University, 
Wilberforce Road, Cambridge CB3 0WA, UK
}
\address{$^c$ Department of Physics and Astronomy, University of Rochester, Rochester
NY~14627-0171, USA
}
\address{$^d$ Department of Physics, University of Durham, Durham DH1 3LE, UK
}
\address{$^e$ Department of Mathematical Sciences, University of Durham, Durham DH1 3LE, UK
}
\address{$^f$ Theory Division, CERN, CH-1211 Geneva 23, Switzerland
}
\address{$^1$ email: {\em carlo@hep.phy.cam.ac.uk}
}
\address{$^2$ email: {\em orr@urhep.pas.rochester.edu}
}
\address{$^3$ email: {\em W.J.Stirling@durham.ac.uk}
}
\address{$^4$ email: {\em webber@hep.phy.cam.ac.uk}
}


\begin{abstract}
Hadron colliders can provide important tests of BFKL `small-$x$' 
dynamics. We discuss two examples of such tests, the inclusive
dijet jet cross section at large rapidity separation and the number
of associated `mini-jets' in Higgs boson production.
\end{abstract}




\section{Introduction}

There has been considerable interest in recent years in QCD scattering processes
in the so-called `high-energy limit', i.e. processes in which
$s \gg |t| \gg \Lambda_{\rm QCD}$. The corresponding cross sections are
controlled by BFKL dynamics \cite{Fadin:1975cb,Balitsky:1978ic}, 
in which  large $\ln(s/|t|)$ logarithms 
arising from soft and virtual gluon emissions are resummed to all orders in
perturbation theory. In the leading logarithm approximation, the energy dependence
of the cross section is controlled by the (hard) BFKL pomeron:
$\sigma \sim s^\lambda$ with $\lambda = \alpha_s 12 \ln 2/\pi$. 

The paradigm BFKL process is deep inelastic scattering at small Bjorken $x$,
for which $t \sim -Q^2$, $s \sim Q^2/x$.  Resummation of the leading $\alpha_s
\ln(1/x)$ logarithms leads to the characteristic $F_2 \sim x^{-\lambda}$ behaviour for
the structure function as $x \to 0$. However it has proved difficult 
in practice to disentangle
BFKL and ordinary DGLAP physics at currently accessible $x$ and $Q^2$ values. One is then
led to consider whether hadron colliders such as the Tevatron and LHC
 can offer a more definitive test of BFKL 
small-$x$ dynamics.

It was first pointed out by Mueller and Navelet \cite{Mueller:1987}
that  production of jet pairs with modest transverse momentum $p_T$ and 
large rapidity separation $\Delta y$ at hadron colliders would
be a particularly clean environment in which to study BFKL dynamics. 
At asymptotic separations the subprocess cross section is predicted
to increase as $\hat{\sigma}_{jj} \sim \exp{(\lambda \Delta y)}$.

To understand the special features of BFKL dynamics, it
will be essential not only to study such fully inclusive cross sections,
 but also to investigate the structure
of the associated final states. For the large $\Delta y$ dijet cross section,
for example, one expects an increasingly large number of `mini-jets',
with transverse momenta  of order $p_T$, produced in
the central region. More generally, one can use BFKL dynamics to predict the expected 
number of such mini-jets in {\it any} small-$x$ hard scattering process at hadron colliders.

In this note we discuss two tests of BFKL dynamics at hadron colliders: the inclusive
dijet cross section and the associated multiplicity of mini-jets in Higgs production.

\section{Dijet cross sections at large rapidity separation} 
\def\pti{p_{T1}}
\def\ptii{p_{T2}}
\def\df{\Delta \phi}
\def\half{{\textstyle{1\over 2}}}
\def\quarter{{\textstyle{1\over 4}}}
We wish to describe events in hadron collisions 
containing two jets with relatively small
transverse momenta $\pti, \ptii$  and large rapidity separation
$\Delta y\equiv y_1 - y_2$. Defining $\df\equiv \vert \phi_1-\phi_2 \vert -\pi$
to be the relative azimuthal angle between the jets, the leading-logarithm 
BFKL prediction for the ($gg$) subprocess
cross section integrated over $\pti, \ptii > p_T$
is  
\beq
{d \hat{\sigma}_{gg} \over  d\df}\Bigg\vert_{\pti^2,\ptii^2 > p_T^2}
 = { 9 \alpha_s^2  \pi\over  2 p_T^2} \;
\frac{1}{2\pi} \sum_{n=-\infty}^{+\infty} \exp{in\df} C_n(t)\; , 
\label{eq:a3}
\eeq
with $t =  3 \alpha_s \Delta y  /\pi$ and 
\bea 
C_n(t) & =& \frac{1}{2\pi}  \int_{-\infty}^{+\infty}
{ dz \over z^2 + \quarter}
\;  \exp{\left(2t \chi_n(z)\right)} \; ,\nonumber \\ 
\chi_n(z) &=& {\rm Re}\,\left[ \psi(1) - \psi\left(\half(1+\vert n\vert)
+iz\right) \right] \; ,
\label{eq:a4}
\eea
where $\psi$ is the digamma function.
The total subprocess cross section, and its corresponding asymptotic behaviour,
is then \cite{Mueller:1987}
\beq
\fl  \hat{\sigma}_{gg}  =  {9 \alpha_s^2  \pi\over  2 p_T^2}\;
 C_0(t)\; , \qquad 
 C_0(t)\ \left\{
\begin{array}{ll}
= 1 & \mbox{for} \ \   t = 0 \\
  \sim  
\left[ \half \pi 7 \zeta(3) t \right]^{-1/2} \; \exp({4 \ln 2\; t})\;
 & \mbox{for} \ \   t \to \infty
\end{array} \right.
\label{eq:a6}
\eeq
from which we see the characteristic BFKL prediction of an exponential
increase in the cross section with large $\Delta y$.
It can also be seen from (\ref{eq:a3}) that the average  
cosine of  the  azimuthal angle 
difference $\df$ defined above is proportional to $C_1(t)$.  In fact we have
\beq
\langle \cos\df \rangle= {{C_1(t)}\over{C_0(t)}}
\label{eq:a7}
\eeq
and as we shall see below, this falls off with increasing $t$.
\begin{figure}
\begin{center}
\mbox{\epsfig{figure=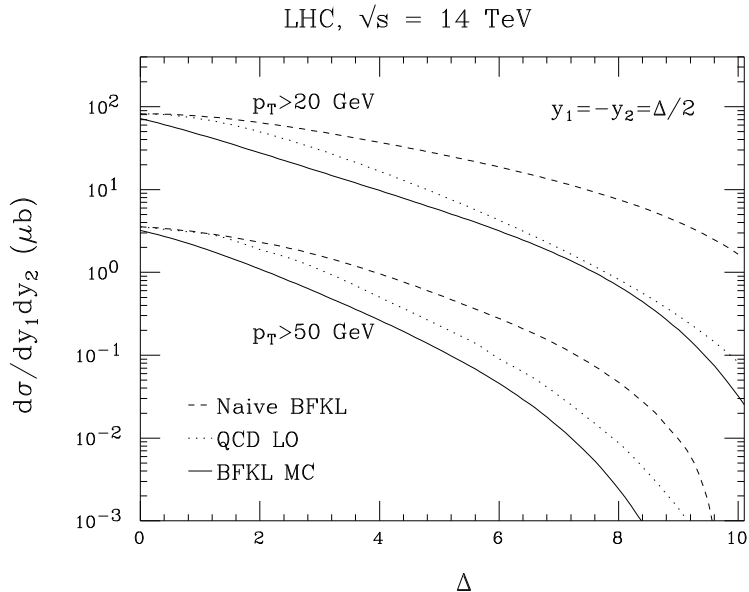,width=12cm}}
\end{center}
\caption{BFKL and asymptotic QCD leading-order 
dijet production cross sections at the LHC ($\sqrt{s}=14$~TeV) 
as a function of 
the dijet rapidity separation $\Delta \equiv \Delta y$.  
The three curves at each transverse momentum 
 threshold 
 use: (i) improved BFKL MC with running $\alpha_s$ (solid lines), 
 (ii) leading-order  BFKL (dashed lines),
  and (iii) the asymptotic ($\Delta y \gg 1$) form
 of QCD leading order (dotted lines).}\label{fig_bfklsig}
\end{figure}

Unfortunately the increase of $\hat{\sigma}$ with 
$\Delta y$ disappears when the subprocess 
cross section is convoluted with parton distribution functions (pdfs), which
decrease with $\Delta y$ more rapidly than $\hat\sigma$ increases. 
This is illustrated in fig.~\ref{fig_bfklsig}. The subprocess cross section
rise at large $\Delta y$ becomes a shoulder in the hadron-level cross section,
whose exact shape depends on the (large-$x$) form of the pdfs.
To avoid this pdf sensitivity, one  can study \cite{DelDuca:1994,Stirling:1994}
the  {\it decorrelation} in $\df$ that arises from
emission of gluons between the jets; BFKL predicts (see eq.~(\ref{eq:a7}))
a stronger decorrelation
than does fixed-order QCD, and this prediction should be relatively 
insensitive to the pdfs.  

In practice it is not useful to compare
analytic asymptotic BFKL predictions  directly
with experiment because nonleading corrections can be large.  In particular,
in the analytic BFKL calculation that leads to (\ref{eq:a3},\ref{eq:a4}) above,
gluons can be emitted arbitrarily, with no kinematic
cost, and energy and momentum are not conserved.  In Ref.~\cite{Orr:1997}
(see also \cite{Schmidt:1997}) a Monte Carlo 
approach is used 
in which the emitted gluons are subject to kinematic 
constraints (i.e. overall energy and momentum are conserved), and other
nonleading effects such as the running of $\alpha_s$ are included.
Kinematic constraints are seen to have a significant effect, suppressing
the emission of large numbers of energetic gluons.  The effect is clearly visible
in fig.~\ref{fig_bfklsig} (solid lines) \cite{Orr:1998}, 
where the `improved' BFKL calculation 
actually gives a {\it smaller} cross section than that at lowest order.
This is due to the sizeable increase in $\hat{s}$, and hence in the large $\Delta y$ pdf
suppression, due to the emitted BFKL gluons.
\begin{figure}
\begin{center}
\mbox{\epsfig{figure=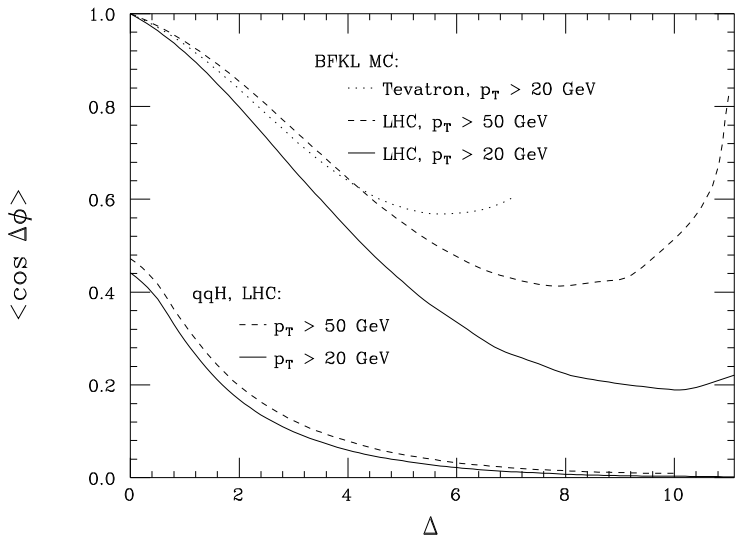,width=12cm}}
\end{center}
\caption{The azimuthal angle decorrelation in dijet production at the Tevatron 
($\sqrt{s}=1.8$~GeV) and LHC ($\sqrt{s}=14$~TeV)
as a function of dijet rapidity difference $\Delta y$.  
The upper curves are computed using the improved BFKL MC with running $\alpha_s$;
they are: (i) Tevatron, $p_T>20$~GeV (dotted curve),
(ii) LHC, $p_T>20$~GeV (solid curve), and (iii) LHC, $p_T>50$~GeV
(dashed curve).  The lower curves are for dijet production in the process
$qq\to qqH$ for $p_T>20$~GeV (solid curve) and $p_T>50$~GeV
(dashed curve).}\label{fig_bfklphi}
\end{figure}

The azimuthal decorrelation is also  weaker in the more realistic 
BFKL calculation.
This is illustrated in fig.~\ref{fig_bfklphi}, where we show \cite{Orr:1998}
the mean value of $\cos{\Delta\phi}$ in dijet production in the improved BFKL
MC approach (upper curves).
The jets are completely correlated (i.e. 
back-to-back in the transverse plane) at $\Delta y  =0$, and as $\Delta y$
increases
we see the characteristic BFKL decorrelation, 
followed by a flattening out and then an increase in 
$\langle\cos{\Delta\phi}\rangle$
as  the kinematic limit is approached.  Not surprisingly, the kinematic
constraints have a much stronger effect when the $p_T$ threshold is 
set at $50$~GeV (dashed curve) than at $20$~GeV (solid curve); 
in the latter case
more phase space is available to radiate gluons.  We also show for 
comparison the decorrelation for dijet production at the Tevatron
for $p_T>20$~GeV.  There we see that the lower collision energy (1.8~TeV)
limits the allowed rapidity difference and 
substantially suppresses the decorrelation at large $\Delta y$.  Recent measurements
of the dijet decorrelation by the D0 collaboration \cite{Dzero:1996} 
at the Tevatron  are in reasonable agreement
with the improved BFKL parton-level predictions. Note that the larger
centre-of-mass energy compared to  transverse momentum threshold 
at the LHC would seem to give it a significant advantage as far 
as observing BFKL effects is concerned.

The lower set of curves in fig.~\ref{fig_bfklphi} refer to Higgs production
via the $WW,\; ZZ$ fusion process $qq \to qq H$, and are included for comparison
\cite{Orr:1998}.
This process automatically provides a `BFKL-like' dijet sample with large
rapidity separation, although evidently the jets are significantly less correlated
in azimuthal angle.

\section{Associated Jet Multiplicities in Higgs Production at the LHC} 

One important aspect of the final state at the LHC 
is the number of mini-jets produced. By mini-jets we mean jets with
transverse momenta above some resolution scale $\mR$ which is very much
smaller than the hard process scale $Q$.  Then the mini-jet multiplicity at
small $x$ involves not only $\ln x\gg 1$ but also another large logarithm,
$T=\ln(Q^2/\mR^2)$, which needs to be resummed.  The results
derived in \cite{Ewerz:1999fn,Ewerz:1999tt} include all terms
of the form $(\as\ln x)^n T^m$ where $1\le m\le n$.
Terms with $m=n$ are called double-logarithmic (DL)
while those with $1\le m<n$ give single-logarithmic (SL) corrections.
In the calculations the BFKL formalism \cite{Fadin:1975cb,Balitsky:1978ic} 
has been used, but the results are expected to hold \cite{Salam:1999ft}
also in the CCFM 
formalism \cite{Ciafaloni:1988ur,Catani:1990yc,Catani:1990sg,Catani:1991gu} 
based on angular ordering of gluon emissions.
 
In order to find $\VEV{r}(x)$, the mean number of resolved mini-jets
as a function of $x$, it is convenient to compute first the Mellin transform
of this quantity
\begin{equation}\label{mellin}
\VEV{r}_\om = \int_0^1 dx\,x^\om\,\VEV{r}(x)\;.
\end{equation}
We find \cite{Ewerz:1999tt}
\begin{equation}\label{meanres}
\VEV{r}_\om = 
- \frac{1}{\chi'} 
\left(\frac{1}{\glip} + \frac{\chi''}{2 \chi'}+ \chi \right) T
- \frac{1}{2 \chi'} T^2
\end{equation}
where $\glip$ is the Lipatov anomalous dimension which solves 
\begin{equation}\label{omlip}
\omega = -\alb\left[2\gE+\psi(\gamma)+\psi(1-\gamma)\right]
\equiv \alb\,\chi(\gamma)\;.
\end{equation}
Here $\alb= 3\as/\pi$, $\psi$ is the digamma function and 
$\gE$ the Euler constant.  In eq.~(\ref{meanres}),
$\chi'$ means the derivative of $\chi(\gamma)$ evaluated at
$\gamma=\glip$. The corresponding expression for the variance 
in the number of jets,
$\sig^2_\om\equiv\VEV{r^2}_\om -\VEV{r}^2_\om$,
is more complicated, see \cite{Ewerz:1999tt}.

To invert the Mellin transform (\ref{mellin}), we can expand
eq.~(\ref{meanres}) perturbatively as a series in $\alb/\om$
and then invert term by term using 
\begin{equation}\label{melom}
\frac{1}{2\pi i} \int_{\half-i\infty}^{\half+i\infty}
d\om\,x^{-\om-1}\left(\alom\right)^n
=\frac{\alb}{x}\frac{[\alb\ln(1/x)]^{n-1}}{(n-1)!}\;.
\end{equation}
The factorial in the denominator makes the resulting series 
in $x$-space converge very rapidly. It is then straightforward to compute the 
mini-jet multiplicity associated with pointlike scattering on
the gluonic component of the proton at small $x$ 
using 
\begin{equation} \label{nx}
n(x)=\frac{F(x,Q^2) \otimes \VEV{r}(x)}{F(x,Q^2)}
\end{equation}
where $F(x,Q^2)$ is the gluon structure function and $\otimes$ represents
a convolution in $x$.

\begin{figure}
\begin{center}
\input{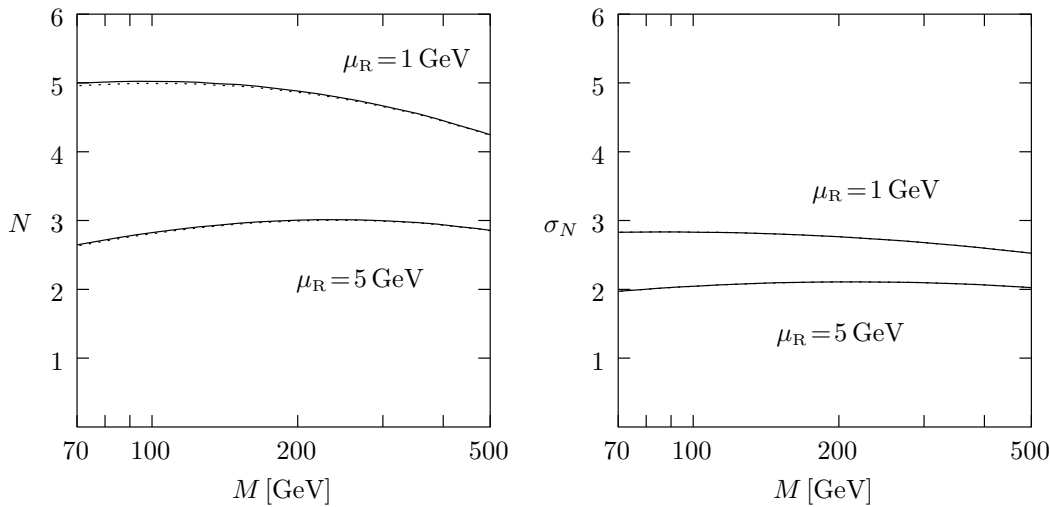}
\end{center}
\caption{The mean value and dispersion of the number of (mini-)jets
    in central Higgs production 
    at LHC for two different resolution scales $\mR$. Solid lines 
    show the SL results up to the 15th order in perturbation theory, 
    dashed lines correspond to the DL approximation.}\label{fig_higgsjets}
\end{figure}

As an application of these results, 
we can compute the 
mean value $N$ and the dispersion $\sig_N$  of the associated
(mini-)jet multiplicity in Higgs boson production at the LHC,
assuming the dominant production mechanism to be gluon-gluon
fusion. At zero rapidity we have gluon momentum fractions
$x_1=x_2=x=M_H/\sqrt s$ where $M_H$ is the Higgs mass,
and $N=n(x_1)+n(x_2) = 2n(x)$.  Similarly $\sigma^2_N(x)
= \sigma^2_n(x_1)+\sigma^2_n(x_2) = 2 \sigma^2_n(x)$.

The results are shown in fig.~\ref{fig_higgsjets}. We see that in
this case the DL results give an excellent approximation and 
the SL terms are less significant.
The mini-jet multiplicity and its dispersion are
rather insensitive to the Higgs mass at the energy of the LHC.
The mean number of associated mini-jets is fairly low, such 
that the identification of the Higgs boson should not be seriously 
affected by them.

\end{document}